\documentclass[12pt,a4paper]{article}

 \usepackage[dvips]{graphics}
 \usepackage{graphicx}
 \usepackage{afterpage}
 \usepackage{enumerate}
\begin{document}
 %%==========================================================

\title{\bf  Two-dimensional MHD models of solar magnetogranulation.
Testing of the models \\ and methods of Stokes diagnostics
  }

 \author{\bf V.A. Sheminova }
 \date{}

 \maketitle
 \thanks{}
\begin{center}
{Main Astronomical Observatory, National Academy of Sciences of Ukraine
\\ Zabolotnoho 27, 03689 Kyiv, Ukraine \\}
\vspace{0.5cm}

\end{center}

 \begin{abstract}
We carried out the Stokes diagnostics of new two-dimensional magnetohydrodynamic
models with a continuous evolution of magnetogranulation in the course of two hours
of the hydrodynamic (solar) time. Our results agree satisfactorily with the results
of Stokes diagnostics of the solar small-scale flux tubes observed in quiet network
elements and active plages. The straightforward methods often used in the Stokes
diagnostics of solar small-scale  magnetic elements were tested by means of the
magnetohydrodynamic models. We conclude that the most reliable methods are the
determination of magnetic field strength from the separation of the peaks in the
Stokes V profiles of the infrared Fe I line 1564.8~nm and the determination of the
magnetic inclination angle from the ratio $\tan^2 \gamma \approx (Q^2 +
U^2)^{1/2}/V^2$. The lower limits for such determinations are about 20~mT and
$10^{\circ}$░, respectively. We also conclude that the 2D MHD models of solar 
magnetogranulation are in accord with observations and may be successfully used to study
magnetoconvection in the solar photosphere.
 \end{abstract}
%-------------------------------------------------

\section{Introduction}

Solar small-scale flux tubes still remain spatially unresolved. In this case the
Stokes diagnostics is the only available technique for the investigation of their
structure and dynamics. The straightforward methods of Stokes diagnostics
\cite{16,19,20} allow the necessary information to be obtained after simple
calculations from direct measurements of the observable Stokes profile parameters.
The inversion methods \cite{4,14} are used to construct flux tube models by fitting
synthetic Stokes profiles to the observed ones. The methods of numerical
calculation of magnetoconvection in the photospheric layers \cite{1,6,7} and
construction of self-consistent nonstationary magnetohydrodynamic models have also
been developed. The MHD models are very useful in studying physical processes in
solar magnetic features, but they cannot be directly compared to observations.

We attempted to make such comparison, using the Stokes diagnostics. In this paper
we match new two-dimensional MHD models \cite{3} to observations and examine the
reliability of the Stokes diagnostics based on the MHD models.

\section{Calculations
}

The MHD models used in our study are described in detail in \cite{3}. The model
sequence starts with a convective model with an initial average magnetic field of
about 5~mT and terminates at the moment 120 min with an average field of 50~mT. The
sequence contains 94 2-D models with a 1-min interval and 52 models with a 0.5-min
interval between them. The simulation region is a rectangle of length $x = 3920$~km
and height $h = 1820$~km, it contains 112 vertical columns (rays) spaced at 35-km
intervals. All the atmospheric thermodynamic parameters necessary for calculating
the equations of radiative transfer in spectral lines for every such column in the
presence of magnetic fields were put at our disposal by A. S. Gadun.

The transfer equations for polarized radiation in each of 112 columns were solved
in the LTE approximation for a plane-parallel atmosphere, and the Stokes profiles
obtained in the solution were averaged over space. The equations were solved
numerically by a modified method \cite{11} which is described in detail by
Sheminova \cite{24}. The Stokes profiles were calculated for three Fe I lines in
the visible ($\lambda\lambda$ 524.71, 525.02, 525.06) and one line in IR ranges
($\lambda~1564.85$~nm).

\section{Comparison of MHD   models with observations
}

The validity of any numerical simulation can be confirmed only by observation data.
The MHD models used here may be quite adequate in their parameters to solar
small-scale magnetic elements. We can calculate the Stokes profile parameters for
these models and compare them to the data of spectropolarimetric observations. For
this purpose we use the Stokes profiles observed in quiet network elements and
active plages with a Fourier spectrograph at the McMath telescope in 1979
\cite{23}. The chief value of these data is that a large number of spectral lines
were observed simultaneously in the wavelength range 445--557~nm with a high
spectral resolution (420000). Unfortunately, the spatial and temporal resolutions
were not high ($10^{\prime\prime}$ and 35--52 min). Out of 402 iron lines, we
selected 170 unblended lines with known laboratory wavelengths. We calculated
absolute shifts for the I and V profiles of these lines, using the strong Mg I
517.2~nm line as a reference line and the laboratory wavelength system by Pierce
and Breckinridge \cite{13}. The magnetic field strength was calculated for every
line by the method of the center of masses \cite{15}. We also determined the
asymmetry of amplitudes and areas in the observed V profiles. Thus we acquired the
necessary observational data which could be used to check the results of
theoretical simulations of magnetoconvection in the photospheric layers.

The time series of MHD models selected for comparison with observations extended
over the simulation period from 95 min to 120 min. We calculated the Stokes
profiles of four spectral lines for each column in the simulation region in each of
52 models (at 0.5-min intervals); we obtained about 5800 profiles for every line.
This procedure is analogous to an observation of the Stokes profiles by scanning a
very narrow region 3920~km long on the solar surface with a spatial resolution of
35~km at 0.5-min intervals in the course of 25 min with instantaneous exposures.
For each calculated profile we determined the same parameters as for the observed
profiles. A preliminary analysis revealed that the profiles of the weak IR line
1564.8~nm were the best suited for the comparison. This line forms deep in the
photosphere ($\log \tau_R$ from 0 to -1), it is the least sensitive to temperature
fluctuations \cite{19} and is virtually unaffected by saturation. The other three
lines, in the visible range, are formed high in the photosphere ($\log \tau_R$ from
-2 to -4), their central intensities are temperature-sensitive, they are also
sensitive to saturation and NLTE effects. The data for the uppermost layers, which
depend on boundary conditions to a greater extent, are less reliable in our MHD
models. The visible lines formed in the upper layers are supersaturated, as a rule,
and their diagnostic capabilities are drastically impaired.

%------------------------------------------------------------- Fig1
\begin{figure}[t]
   \centering
   \includegraphics[width=10.5cm]{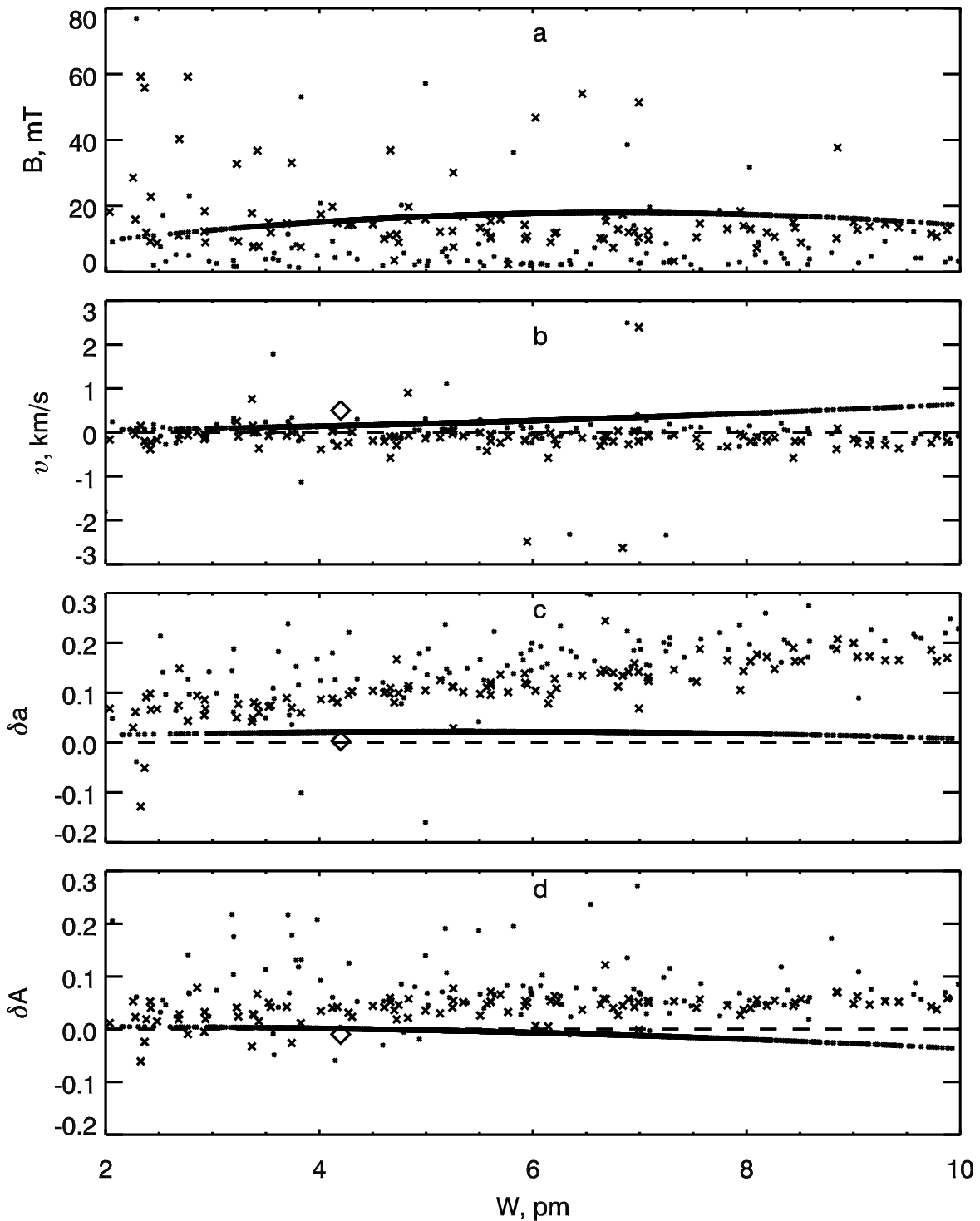}
 \hfill
\parbox[b]{5.1 cm}{ \vspace{0.0cm}
%   \caption[]
{Fig.~1.
Comparison of the results of Stokes diagnostics of MHD models (solid line) and
observations in the quiet network features (squares) and active plages (crosses):
a) field intensity, b) absolute shift, c) asymmetry of V-profile amplitudes, d)
asymmetry of V-profile areas.
 } \label{Fig:Fig1}
   }

\end{figure}
%_____________________________________________________________

Figure 1 displays the magnetic field strength $B$, absolute shift $v$, asymmetry of
amplitudes and areas, $\delta a$ and $\delta A$, as functions of equivalent width;
the quantities obtained from the observed V profiles are shown by squares and
crosses, and the calculated parameters are represented by the second-order curves
(average values from 5800 IR line profiles). We did not expect a complete agreement
between the observed and calculated parameters, since there were some distinctions
in resolution, line sampling, and selection of observation regions. The observed
Stokes profiles were measured in magnetic features only, while the calculated
profiles referred to the entire simulation region. Only 30 percent of calculated
scans, in the best case, may be referred to magnetic elements. It is seen from
Fig.~1 that the calculated values of $B$ and $v$ fall within the observed intervals,
while the asymmetries of amplitudes and areas are smaller, on the average, than in
the observed profiles. This might be attributed to the well-known observation facts
\cite{12} . In the quiet network elements the measured asymmetries  $\delta a$  and
$\delta A$ for 1564.8~nm are close to zero (big diamonds in Fig.~1). An
interpretation for these facts can be found in \cite{7}. Despite some differences,
we may conclude that realistic Stokes profile parameters are derived within the
framework of the MHD models, they are close to the parameters observed in the quiet
network elements and plages.

\section{Reliability of Stokes diagnostics
}

To test the methods of Stokes diagnostics, we selected one 2-D model (snapshot) from the MHD model sequence  \cite{3}, namely, the model corresponding to the 120th min of simulation time.

Figure 2 demonstrates the distribution of various parameters in the
simulation region: isotherms, isobars, velocity field, and field strength, together with the field lines and polarity. Two flux tubes of various strengths and polarities with clearly defined strong longitudinal fields stand out in the simulation region; an area at the granule center, where a new flux tube begins to form, can be also clearly seen. With these data, we calculated the Stokes profiles of four spectral lines for each of 112 model columns (scans) and applied various methods of Stokes diagnostics with the aim to determine the flux tube parameters. We examine the agreement between the quantities thus obtained and the thermodynamic model parameters and the observation data for magnetic elements.

%------------------------------------------------------------- Fig2
\begin{figure}
   \centering
   \includegraphics[width=13.5cm]{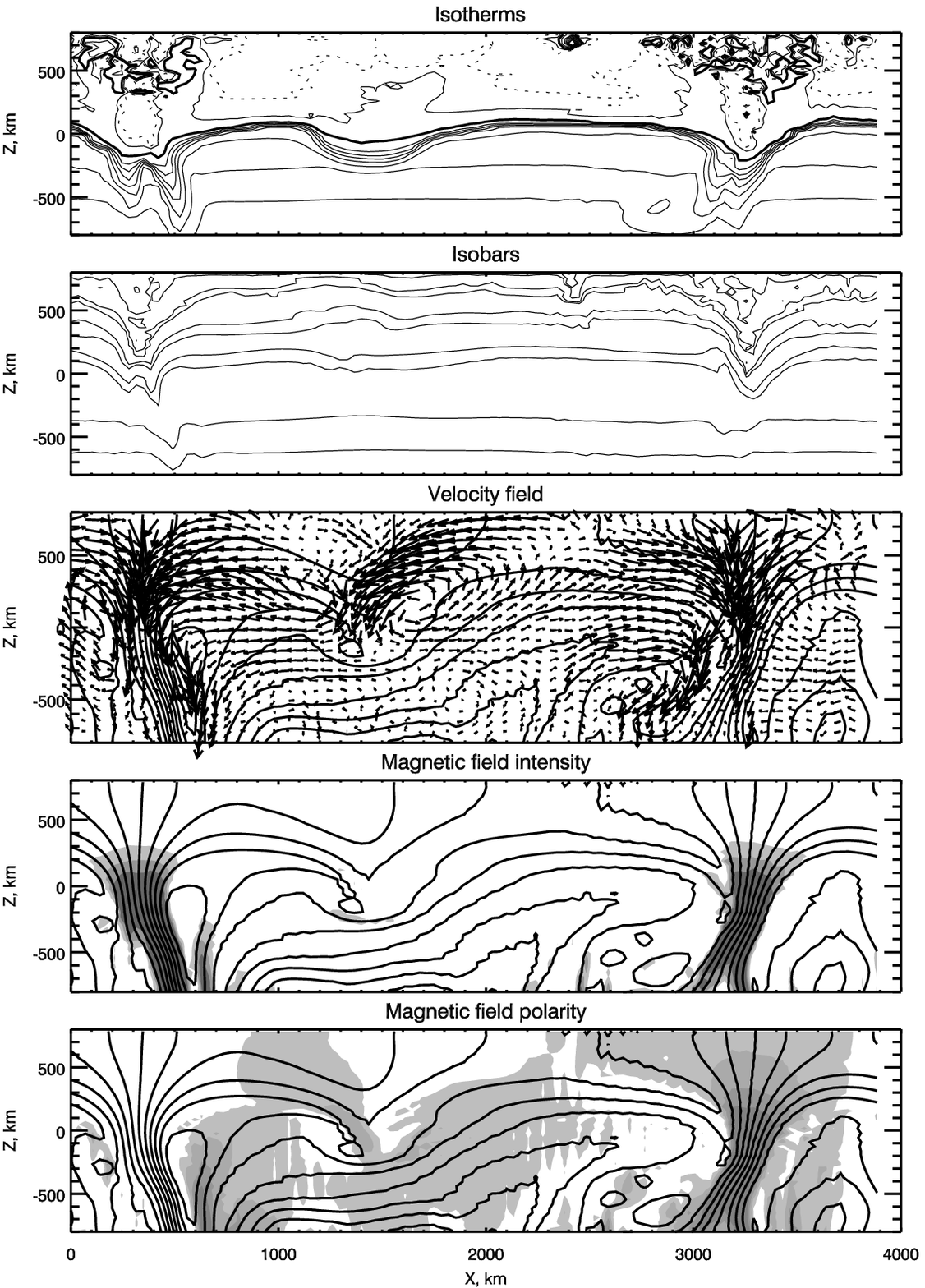}
% \hfill
\parbox[b]{15.cm}{ \vspace{0.5cm}

%   \caption[]
{Fig.~2. Snapshots from 2D MHD models  \cite{3}  
at simulation moment 120 min. Isotherms for every 1000~K: dotted line) 4000 K and thick
line) 6000 K. The latter roughly indicates the level $\log \tau_R = 0$. Magnetic
field lines are shown in three lower panels. Magnetic field intensity: hatching
density is proportional to intensities of 0, 80, 110, and 140~mT. Magnetic field
polarity: hatching density corresponds to field intensities of 1, 40, 80, and
120~mT in the positive polarity field.
 }
      \label{Fig:Fig2}
}
\end{figure}
%_____________________________________________________________
%------------------------------------------------------------- Fig3
\begin{figure}[t]
   \centering
   \includegraphics[width=9.cm]{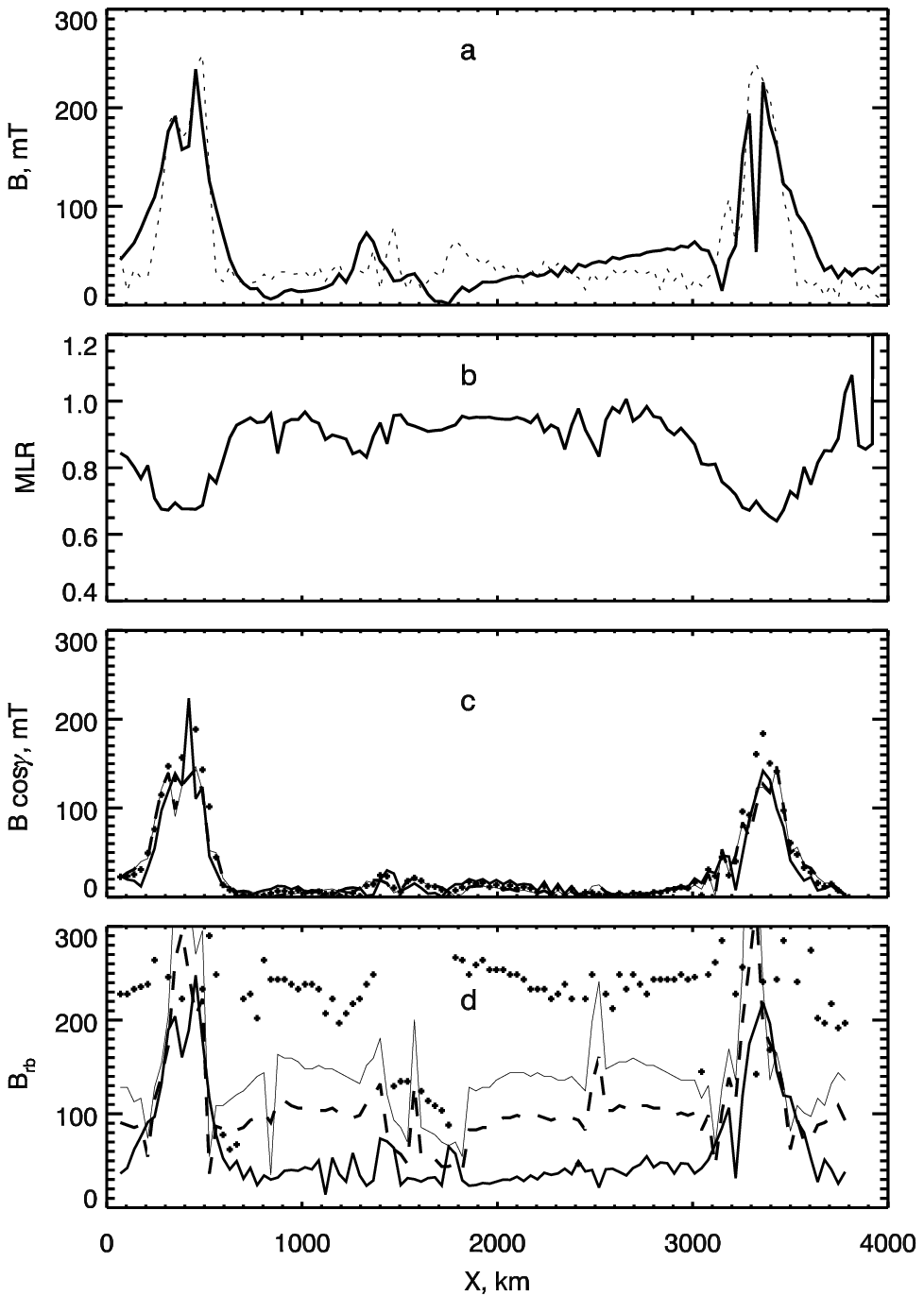}
 \hfill
\parbox[b]{6.4cm}{ \vspace{0.0cm}

%   \caption[]
{Fig.~3. 
Magnetic field intensity $B$ along the simulation region: a) from a MHD snapshot
at the levels  $\log \tau_R = 0$ (solid line) and  $\log \tau_R = -2$ (dotted
line); b) line ratio method; c) method of center of masses; d) method based on the
distance between the V-profile peaks. For c) and d) thick line shows the date
derived from $\lambda$ 1564.85~nm, thin line from $\lambda$  524.71~nm, dashed line
from $\lambda$~525.02~nm, and plusses from $\lambda$ 525.06~nm.
 }
      \label{Fig:Fig2}
}
\end{figure}
%_____________________________________________________________

\subsection{Magnetic field strength
}

There are three methods for the determination of this parameter. The first method
is based on the ratio of the amplitudes $a_V$ of the V profiles of two lines in the
visible range with various Zeeman splittings and similar other parameters
\cite{21,22}:

\[{\rm MLR}= \frac{ a_V(525.02)g_{\rm eff}(524.71)} {a_V(524.71)g_{\rm eff}(525.02)} \approx
\frac{2}{3}\cdot \frac{a_V(525.02)}{a_V(524.71)}.\]

Traditionally, the observed magnetic line ratio (MLR) is calibrated with the use of
simple models with homogeneous magnetic fields; MLR~$=1$ for weak fields, and MLR~$ <
1$ for strong fields. This ratio may be greater than unity in the regions where the
saturation in the cores of strong lines weakens the V profile. When the velocity,
which broadens the profiles, increases, the MLR may also increase. This degrades
the reliability of the field strength determined for flux tubes and makes more
stringent the requirements imposed on the MLR calibration.

In the method of the centers of mass the longitudinal field strength is estimated
from the measurements of the central wavelengths of positively and negatively
polarized line components \cite{15}.

\[\lambda_{\pm}= \frac{\int [I_c-(I_c\pm V)]\lambda d\lambda} {\int [I_c-(I_c\pm V)] d\lambda}  \]

The field strength $B$ (in tesla) is readily found from the well-known expression
for the Zeeman splitting:

\begin{equation}
B_{rb}=(\lambda_{-}-\lambda_{+})/(2\cdot 4.67\cdot 10^{-8}\lambda_{0}^2g_{\rm
eff}),
\end{equation}
 $\lambda_{0}$  being the line wavelength (in nanometers). This method is quite
reliable for the solar disk center, but it gives only the lower limit of $B$, as it
specifies the longitudinal field component only.

The third method is the simplest one --- it is based on the measurement of the
distance $\Delta\lambda_{br}$ between the peaks in the red (r) and blue (b) wings
of the V profile of the IR line with the Land\'{e} factor equal to 3. This is a
very efficient method \cite{9,19}. With expression (1), we can derive the field
strength $B_{br}$ as the upper limit. As the quantity $\Delta\lambda_{br}$ consists
of the Doppler line width $\Delta\lambda_{D}$ and the Zeeman splitting
$\Delta\lambda_{H}$, the distance between the V-profile peaks in weak fields, when
the first component is prevailing, is determined mainly by the nonmagnetic line
width, i.e, it is practically independent of magnetic field. As the field
increases, the Zeeman splitting becomes prevailing. That is the reason why this
method fails in the case of very weak photospheric fields.

One can see in Fig.~3a the variations in the model field strength at the
photosphere levels $\log\tau_R = 0$ and  $\log\tau_R = -2$. The maximum field
strength in the model flux tubes at  $\log\tau_R = 0$ is as large as 200--250~mT in
a very narrow interval (35--70~km). It is known from observations (their resolution
is no better than 200~km) that the field strength at  $\log\tau_R = 0$ is 100--150
mT in magnetic elements and 150--200~mT in pores, less than in the model flux
tubes.

Figures 3b-d show the field strength determined by three methods. The best fit to
the model values of $B$ is provided by the third method (distance between the
V-profile peaks) with the use of the IR line Fe I 1564.8~nm (Fig.~3d). The lower
limit for the measurements by this method is about 20~mT. The line ratio method
(Fig.~3b) becomes less reliable at about 0.1~T, MLR being equal to 0.65--0.7 in
this case. The method of the center of masses (Fig.~3c) allows longitudinal
magnetic fields to be measured, and their strength obviously depends on the
inclination of field lines in flux tubes. A greater inclination in the second tube
as compared to the first tube results in an underestimate of about 70~mT at the
level  $\log\tau_R = 0$.

\subsection{Magnetic field inclination
}

The angle $\gamma$ of the field vector and the line of sight can be found directly
from the relations for the amplitudes of $\sigma$-components in the V, Q, and U
profiles \cite{5,16}:

\[ V \approx \cos\gamma,~~~~Q  \approx \sin^2 \gamma \cos^2 \chi,
~~~~ U \approx \sin^2\gamma \sin^2 \chi ,\]\

 \[\frac{\sqrt{Q^2+U^2}}{V} \approx \frac{\sin^2 \gamma}{\cos \gamma},~~~~ \frac{U}{Q}=\tan^2\chi.\]
 Here $\chi$  is the field vector azimuth. These relations are valid when the observed line
is weak and the angles $\gamma$, $\chi$  do not change along the line of sight. The
ratio $(Q^2 + U^2)^{1/2}/V$ depends on wavelength, field strength, line saturation,
etc. \cite{19}.

In strong magnetic fields the amplitudes of $V$, $Q$, and $U$ profiles do not
depend on $B$ due to magnetic saturation of lines, and therefore $\gamma$ can be
reliably determined. In weak fields, $V \approx~ B$ and Q and U are of the order of
$B^2$, and the ratio depends not only on $\gamma$ but on $B$ as well. In actual
practice this ratio is calibrated with the use of model calculations. In our
two-dimensional MHD models the magnetic vector azimuth is equal to zero or
$180^\circ$ and $U \approx 0$. Although the $U$ profile does not vanish altogether
due to magnetooptical effects, it is very small, and it may be ignored. The angle
$\gamma$ can be derived directly from the amplitude ratio when the ratio $(Q^2 +
U^2)^{1/2}/V$ is divided by $V$ and $U = 0$ is substituted:

\[ \frac{\sqrt{Q^2+U^2}}{V^2}\approx \frac{\sin^2\gamma}{V\cos \gamma}=\tan^2\gamma,
~~~~ \frac{Q}{V^2} \approx \tan^2\gamma , ~~~~\frac{\sqrt{Q}}{V} \approx \tan\gamma
.\]

In this version the ratio is independent of $B$ under the conditions of weak fields
as well as strong ones. The equality of the amplitude ratio and $\tan^2\gamma$ is
only approximate --- the angle being determined is affected by velocities and other
parameters. We tested the accuracy of this approximation by calculating $\gamma$
from the ratios $\sqrt{Q}/V$ for four lines and comparing the calculated values
with the model ones.

%------------------------------------------------------------- Fig4
\begin{figure}[t]
   \centering
   \includegraphics[width=9.cm]{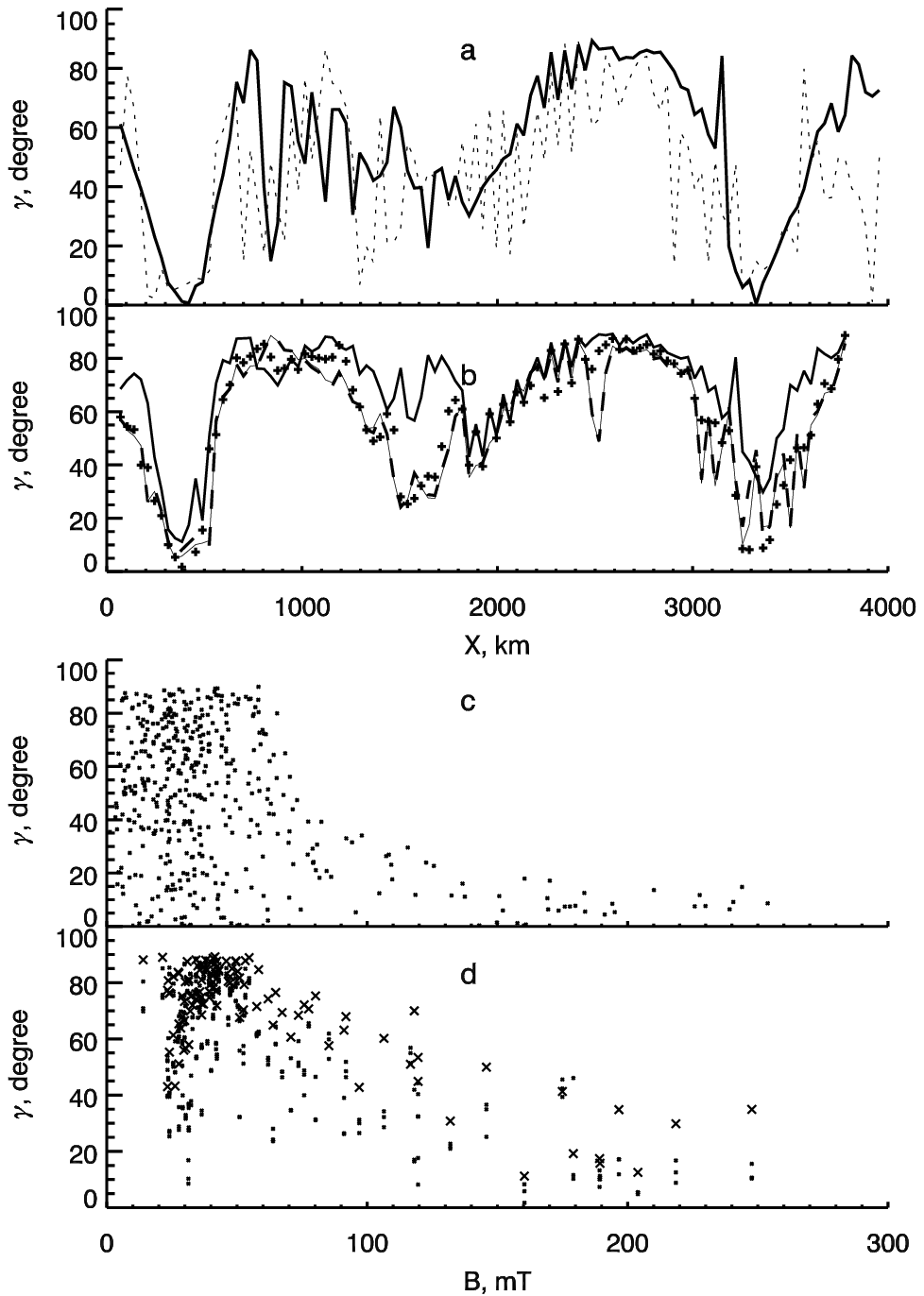}
    \hfill
\parbox[b]{6.4cm}{ \vspace{0.0cm}
   %\caption[]
   {Fig.~4(a, b). 
Magnetic field inclination along the simulation region: a) from  a MHD snapshot at
the levels $\log \tau_R = 0$ (solid line) and $\log \tau_R = -2$ (dotted line), b)
from the amplitude ratio $\sqrt{Q}/V$ for four lines.  Note the data derived from
$\lambda$~524.7~nm (thin line) and $\lambda$~525.02~nm (dashed line) are almost the
same. The remaining symbols are the same as in  Fig.~3c, d.}

\vspace{0.5cm}{Fig.~4(c, d). The inclination $\gamma$ vs. magnetic field
intensity: c) from a MHD snapshot at four $\log \tau_R $ levels --- 0, -1, -2, and -3;
d) $\gamma$ from the amplitude ratio $\sqrt{Q}/V$ for four lines (crosses
correspond to $\lambda$~1564.8~nm). Here $B$ was derived  from the distance between 
the V~peaks of the $\lambda$ 1564.8~nm line.
 }
       \label{Fig:Fig4}
       }
\end{figure}
%_____________________________________________________________

Figures 4a, b show the variations of $\gamma$ along the simulation region, and Figs
4c, d show $\gamma$ as a function of $B$. The inclination is 5--10$^\circ$ within a
range of 300 km in the central part of the first flux tube and 10--20$^\circ$ in
the second flux tube. This is in accord with the data of \cite{18}, where an
analysis of the V, Q profiles in plages gave an inclination of flux tubes no less
than 10$^\circ$. As judged from the distribution $\gamma(B)$ found with the use of
the Stokes diagnostics (Fig.~4d), there are no fields weaker than 20~mT, but such
fields cannot be detected by the method based on the measurements of the V-profile
peaks. It is also obvious from Fig.~4d that there are no longitudinal fields with
$B < 160$~mT and $\gamma< 10^\circ$. This is likely to be a result of the
limitedness of the method used to determine $\gamma$. Besides, the angles $\gamma$
found with the use of the Fe~I~1564.8~nm line are greater than the angles
determined from the lines in the visible range. Hence we may conclude that the
ratio $\sqrt{Q}/V$ for the visible lines gives angles overestimated by
$10^{\circ}$, and the overestimation is even greater for the IR line. The method we
used here is simple and quite reliable, as the distribution $\gamma (B)$ obtained
by it is in agreement with the model distribution. In reality, when $\chi\neq 0$,
we have to use the relation $(Q^2 + U^2)^{1/2}/V^2 \approx \tan^2\gamma$.

\subsection{Radial velocities in magnetic elements
}

As of now, there are no observations which would prove the existence of stationary
vertical flows with velocities higher than 250 m/s (average from numerous
observations) inside small-scale magnetic elements \cite{17}. Considerable line
shifts are sometimes observed in individual magnetic elements. The nature of the
V-profile shifts still remains unknown and is always a topical problem in the
studies of the structure and dynamics of flux tubes. The horizontal distribution of
vertical velocities displayed in Fig.~5a was obtained directly from the MHD models.
A velocity minimum (about 0~km/s) is observed at the center of the first flux tube,
while maxima of 6 and 9~km/s can be seen at a distance of 100~km from the tube
center; then the velocity falls sharply again. Such the distribution suggests the
existence of very sharp horizontal gradients which may be a consequence of a
relatively big step (35~km) in our numerical simulation. In actuality the
velocities are smoother, but just the same, they strongly differ from the
velocities in the nonmagnetic neighborhood. In observations the velocity may be
underestimated because it heavily depends on spatial averaging due to low spatial
resolution and atmospheric and instrumental distortions.

The velocities $v_I,~ v_Q,~ v_V$, which were found from the I, V, and Q profiles
calculated for four lines, were compared with the model velocities (Fig.~5). The
best agreement was found for the velocities measured as shifts of the
$\pi$-components of Q profiles (Fig~5d), except the cases when $\gamma = 0$, $Q =
0$ (flux tube center), and shifts of the zero crossing of V profiles, except the
cases of multicomponent V profiles (Fig~5c). The velocities obtained from the
shifts of the I-profile center for the IR line ((Fig~5b)) deviate strongly from the
true (model) velocities at sites with strong magnetic fields. As the I profile
splits completely into $\sigma$-components of various amplitudes, the risk of
mistaking the wavelength of a strong $\sigma$-component for the wavelength of the
line core is run when the central wavelength of the line is determined
automatically from the central intensity minimum.

So, the vertical velocities inside flux tubes are determined quite reliably from
the shifts of Q and V profiles. The velocity should not be determined in the
automatic mode from the shifts of the I profiles of IR lines with large Land\'{e}
factors.

%------------------------------------------------------------- Fig5

%------------------------------------------------------------- Fig5
\begin{figure}[t]
   \centering
   \includegraphics[width=9.cm]{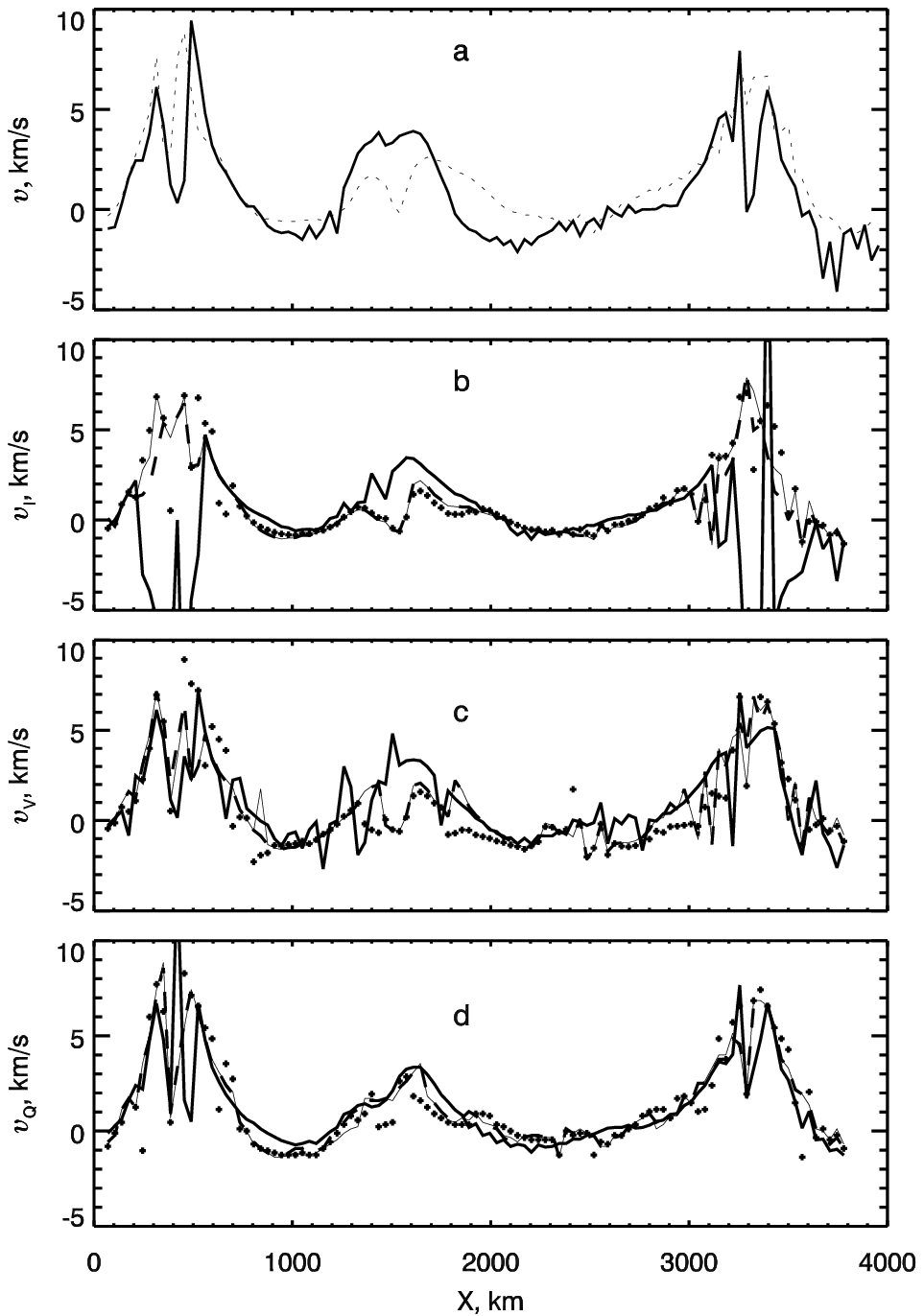}
    \hfill
\parbox[b]{6.7cm}{ \vspace{0.0cm}
%   \caption[]
{Fig.~5.
Vertical velocity along the simulation region: a) from the MHD model at the levels
$\log\tau_R =  0$ (solid line) and $\log\tau_R = -2$ (dotted line); b, c, d) shifts
of the I, V, and Q profiles of four lines, respectively. Designations are the same
as in Fig.~3c, d. }
       \label{Fig:Fig5}
       }
\end{figure}
%_____________________________________________________________

\subsection{Temperature
}

The temperature diagnostics in small-scale magnetic features is carried out with
the use of the temperature line ratio (TLR). This quantity was used for the first
time when the true weakening of the I profile of the line Fe~I~525.02~nm with
respect to the line Fe~I~523.30~nm inside unresolved magnetic elements was studied
\cite{10}. Ideally, the lines in the TLR must have different temperate
sensitivities and close central depths $d_c$, wavelengths, and effective Land\'{e}
factors $g_{\rm eff}$. With the ratio between the amplitudes $a_V$ of their V
profiles, we can estimate the line weakening associated with temperature variations
in a region with magnetic fields, i.e., in a flux tube. At elevated temperatures
the sensitive line is weakened to a greater degree and TLR is less than unity, and
vice versa. To obtain temperature from this ratio, it should be calibrated by model
calculations, as it is done for the magnetic line ratio MLR.

For lines in the visible spectrum, the ratio

\[{\rm  TLR}=\frac{a_V(524.71)g_{\rm eff}(525.06)}{a_V(525.06)g_{\rm eff}(524.71)}
=\frac{1.5}{2} \cdot \frac{a_V(524.71)}{a_V(525.06)}\]
 is often used in actual practice. It does not meet all the requirements,
as the line parameters are not always close to one another, and so we introduced a
coefficient which equalizes the amplitudes of V profiles. This coefficient is
approximately equal to 1.11 on the assumption that $a_V \approx d_c$. We calculated
TLR (with a coefficient of 1.11) for our MHD model (Fig.~6). It depends not only on
temperature, but on the vertical temperature gradients and the magnetic and
temperature saturation in the lines as well, and this distorts the results of the
diagnostics. In the central regions of flux tubes the magnetic saturation of the
lines Fe~I~524.7 and 525.06~nm is very strong. The amplitude $a_V$ is
temperature-insensitive there, and the method fails (TLR~$\approx 1$); TLR~$
> 1$ at flux tube periphery, where the saturation drastically decreases. Hence it
follows that the TLR method for temperature diagnostics inside flux tubes is not
sufficiently reliable because of line saturation effects, especially in those cases
when the spatial resolution is high. The central depths of these two lines become
highly saturated in the regions with lower temperatures and become less sensitive
to temperature variations.

The continuum brightness is also often used as a temperature indicator. Figure 6b
shows the profiles of normalized continuum intensity along the horizontal
coordinate, or the contrast $I_c/{<}I_c{>}$. The intensity $I_c$ was calculated for
each model column in the continuum of two lines Fe~I~525.0 and 1564.8~nm,
${<}I_c{>}$ being the intensity averaged in the horizontal direction over the whole
simulation region. Observations suggest that the contrast in small magnetic
elements heavily depends on spatial resolution. Magnetic elements less than 300~km
in size have a small brightness (1.1--1.4 in the network), and the elements larger
than 300~km have a darker continuum (0.7--0.9 in plages). For IR lines, no
brightening is observed even in small magnetic elements, while darkening is typical
of larger elements. This occurs because the IR lines near $\lambda = 1650.0$~nm
have a minimum in the continuum opacity.

The contrasts calculated for the MHD models at Fe~I~525.0 and 1564.8~nm (dotted
line and solid line, respectively, in Fig.~6b) are very different, the horizontal
gradients being much higher in the visible spectrum. The contrast changes sharply
at the center of the stronger flux tube --- from 0.5 to 1.3, on the average, in the
visible spectrum and from 0.8 to 1.0 in the infrared. The temperature at the level
$\log \tau_R = 0$ changed from 5700 to 6400~K at the center of the same flux tube
(Fig.~6a). Our calculations are in accord with the brightness variations observed
in the continuum in magnetic features.

%------------------------------------------------------------- Fig6
\begin{figure}
   \centering
   \includegraphics[width=9.cm]{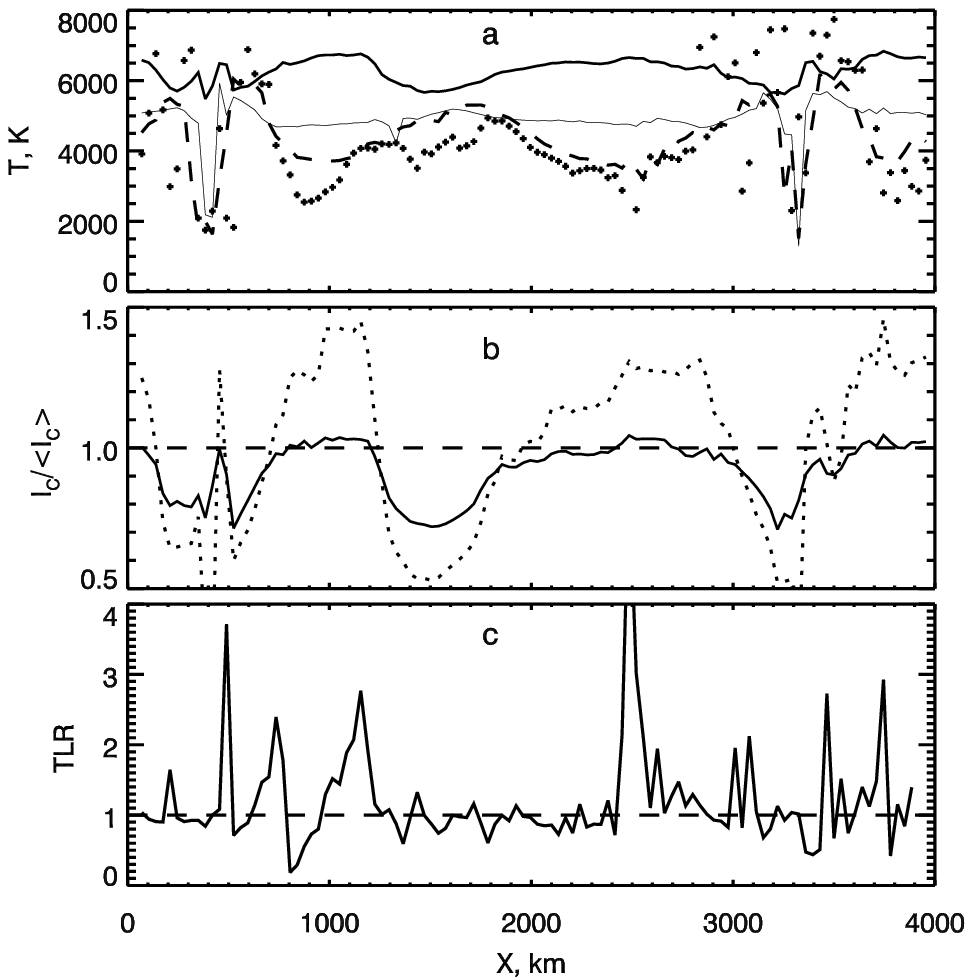}
 \hfill
\parbox[b]{6.5cm}{ \vspace{0.0cm}
%   \caption[]
{Fig.~6. 
Temperature distribution along the simulation region: a) from the MHD model at the
levels $\log \tau_R = 0$ (thick line), $\log \tau_R = -1$ (thin line), $\log \tau_R
= -2$ (dashed  line), and $\log \tau_R = -3$ (plusses); b) continuum contrast at
$\lambda$~525.0~nm (dotted line) and  $\lambda$~1564.8~nm (thick line); c) method
of temperature line ratio. }\label{Fig:Fig6}
 }

\end{figure}
%_____________________________________________________________

\section{Spatial averaging effect in the Stokes diagnostics }

It is well known that not only the amplitudes and asymmetries of the measured
Stokes Q, U, and V profiles but the areas of the profile wings and shifts as well
depend on the spatial, temporal, and spectral resolution. The reason is the spatial
and temporal averaging as well as the instrumental and atmospheric distortions of
the signal measured. Simulations of the spatial smearing made with the use of the
modulation transfer function (which allows for the atmospheric and telescopic
distortions) reveal that the contrast in the continuum of a flux tube 100~km in
diameter may be by a factor of 7--8 higher than the contrast measured with a
0.3$^{\prime\prime}$ resolution. At present the spatial resolution is not better
than 0.25--0.3$^{\prime\prime}$ (180--220km), and the reliability of the parameters
measured in thin flux tubes still remains a problem. Calculating the V profiles for
various intervals along the simulation region in the MHD models, we may study the
spatial averaging effect on these profiles. The averaged profiles may have several
components with widely different parameters depending on horizontal temperature
gradients, vertical velocities, magnetic field intensities, inclinations, and
polarities. The well-known shape of V profiles may be distorted in this case, and
the application of the Stokes diagnostics becomes questionable.

The problem is illustrated in Fig.~7, where we plotted the amplitudes $a_V$ and the
parameters  $B_{rb},~v_V,~ v_Q$ derived from the V and Q profiles of the line
Fe~I~1564.8~nm in spectral scans with varying spatial resolution: about 35, 70,
200, 600, 1000, 1300, and 4000~km. The spatial variations in the velocity and the
magnetic field intensity and polarity demonstrated in Fig.~2 can be compared with
the diagnostics results plotted in Fig.~7. The V profiles will be affected by
spatial resolution until very high resolutions are attained. At a 35-km resolution
the derived values of $B_{rb},~v_V,~ v_Q$ coincide quite well with the model
values. When the V profile consists of two or more components, the distortions of
the profile at low resolutions markedly affect the determination of the amplitude
$a_V = (a_b + a_r)/2$ (Fig.~7a), the distances between the peaks (Fig.~7b), and the
zero-crossing shifts (Fig.~7c). The spatial averaging drastically degrades the
accuracy of the Stokes diagnostics at resolutions lower than 200~km ($\approx
0.3^{\prime\prime}$), especially in the regions with steep horizontal gradients. As
an example, we give the velocities derived from the shifts of the profiles
calculated for the spectral scans with different resolutions from 0 to 1300~km in
the simulation region. We obtained 0.4, -1, 2.4, 0.5, -0.5, and 0~km/s for six
scans with a 200-km resolution, 0.6, -0.2~km/s for two scans with a 600-km
resolution, and 0~km/s for one scan with a 1300-km resolution. The highest velocity
obtained with a resolution of 0.3$^{\prime\prime}$ (200~km) was four times greater
than the velocity obtained with a resolution of 1$^{\prime\prime}$ (600~km). Hence
it follows that the choice of spatial resolution strongly affects the results of
Stokes diagnostics with the use of V profiles.

%------------------------------------------------------------- Fig7
\begin{figure}[t]
   \centering
   \includegraphics[width=9.cm]{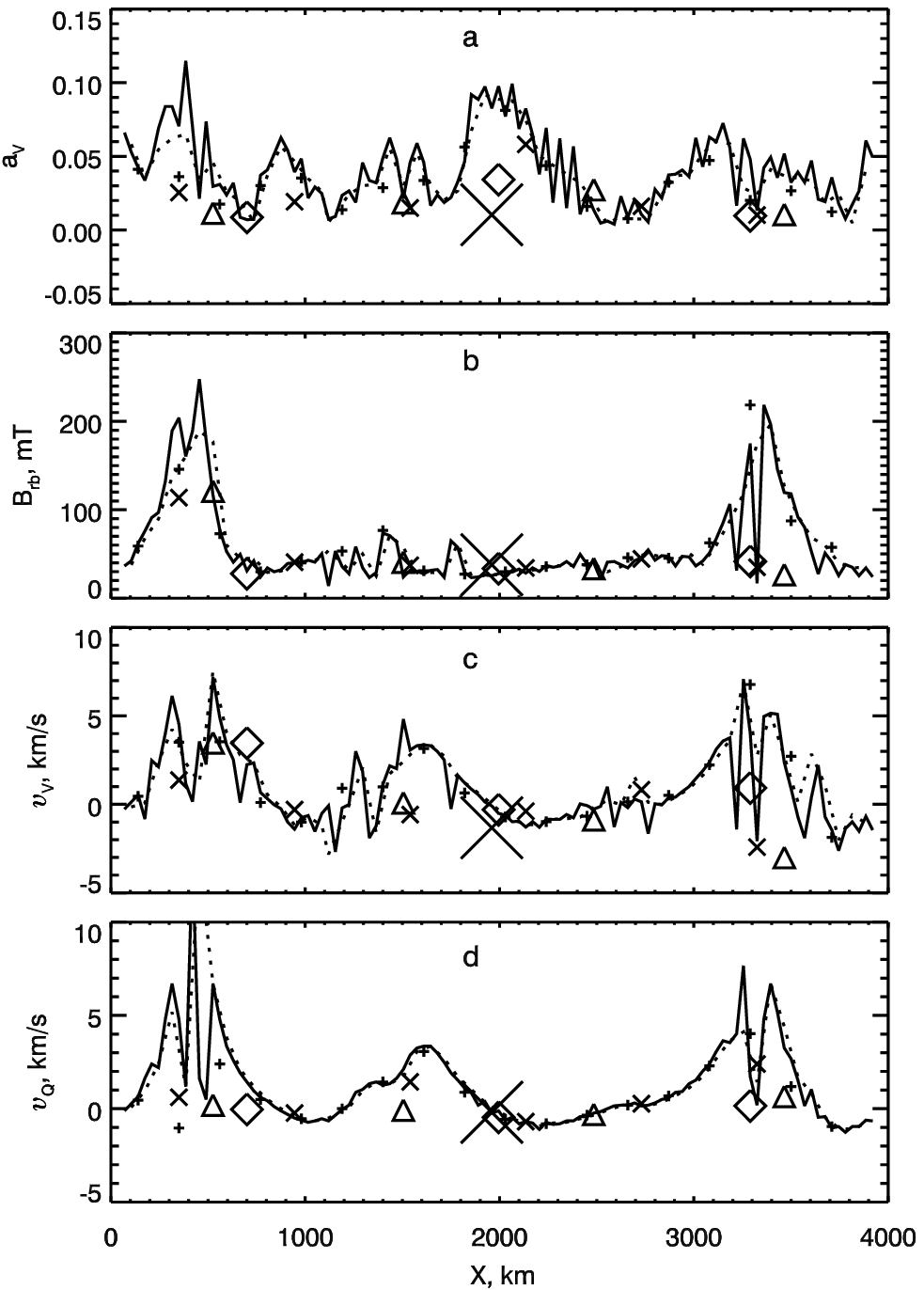}
 \hfill
\parbox[b]{6.5 cm}{ \vspace{0.0cm}
%   \caption[]
{Fig.~7. 
Effect of spatial averaging on the results of Stokes diagnostics, a) amplitude
$a_V$; b) field intensity $B_{rb}$; c, d) shifts of V and Q profiles. Horizontal
averaging of profiles: 35~km (solid line), 70~km (dotted line), 200~km (plusses),
600~km (crosses), 1000~km (triangles), 1300~km (diamonds); the entire region (big
cross).
 }\label{Fig:Fig6}
}

\end{figure}
%_____________________________________________________________

\section{Conclusion }

We demonstrated with the use of the Stokes diagnostics that new 2-D MHD models \cite{3}
are in accord with observations and may be successfully used to study
magnetoconvection in the solar photosphere. We also tested the methods of Stokes
diagnostics of small-scale magnetic elements. The results are as follows.

The IR line $\lambda$~1564.8~nm of Fe I is more suitable for the testing than the
Fe I lines $\lambda\lambda$~524.71 and 525.02~nm. Being less temperature-sensitive,
it is less prone to saturation, and its high magnetic sensitivity does not decrease
inside flux tubes at low temperatures.

The magnetic field intensity is estimated most reliably from the measurements of
the distances between the V-profile peaks in the IR lines with large Land\'{e}
factors. The lower sensitivity limit of this method is about 20~mT. When we want to
determine the longitudinal field intensity only, the method of the center of masses
may be used with confidence for any magnetic lines. Large longitudinal magnetic
fields (~0.1 T) are detected with confidence from the magnetic line ratio (MLR~$ <
0.7$ at the flux tube center).

The inclination of magnetic field vector can be determined directly from the
amplitude ratio $\tan^2\gamma \approx  (Q^2 + U^2)^{1/2}/V^2$ with an accuracy of
10$^\circ$. Longitudinal fields ($\gamma < 10^\circ$, $B < 160$~mT) cannot be
measured by this method.

The velocities of vertical motions inside magnetic elements are reliably determined
from the shifts of the $\pi$-components in the Q profiles and the zero crossings of
V profiles. When the V profiles are of anomalous shape because of the presence of
regions with opposite polarities, the shift of the $\pi$-components of Q profiles
should be used.

The temperature in flux tubes with high spatial resolution cannot be reliably
estimated from the temperature line ratio for two lines in the visible spectrum due
to strong magnetic and temperature saturation of lines in flux tubes.

Spatial averaging significantly affects the Stokes diagnostics of flux tubes.
Resolutions worse than 200~km may produce erroneous results. The reason has to do
with the horizontal gradients of parameters inside flux tubes.

{\bf Acknowledgements.} We are indebted to A. S. Gadun for furnishing the MHD
models for the Stokes profile calculations and S. K. Solanki for furnishing the
observations as well as for useful discussion of the results and their comments.The
study was partially financed by the Swiss National Science Foundation (Grant No.
7UKPJ 48440).

%========-------------------------------------------------

%%%%%%%%%%%%%%%%%%%%%%%%%%%%%%%%%%%%%%%%%%%%%%%%%%%%%%%%%%%%

\end{document}